\documentclass[12pt, draftclsnofoot, onecolumn, romanappendices]{IEEEtran}

\usepackage{amsmath}
\usepackage{amsthm}
\usepackage{amsbsy}
\usepackage{amssymb}
\usepackage{bm}
\usepackage[noadjust]{cite}
\usepackage{graphicx}
\usepackage{float}
\usepackage{epsfig}
\usepackage{latexsym}
\usepackage{wasysym}
\usepackage{placeins}
\usepackage{color}
\usepackage{rotating}
\usepackage[usenames,dvipsnames]{xcolor}
\usepackage{caption}
\usepackage{algorithm}
\usepackage{algpseudocode}
\usepackage{subcaption}
\usepackage{epstopdf}
\usepackage{cancel}
\usepackage{tikz}
\usepackage{tkz-euclide}
\usetkzobj{all}
\usepackage{overpic}

\newtheorem{definition}{Definition}
\newtheorem{lemma}{Lemma}

\DeclareMathAlphabet{\mathbit}{OML}{cmr}{bx}{it}
\DeclareMathAlphabet{\mathsf}{OT1}{cmss}{m}{n}
\DeclareMathAlphabet{\mathTXf}{OT1}{cmss}{bx}{it}

\DeclareMathOperator*{\argmax}{argmax}

\newcommand{\norm}[1]{\lVert{#1}\rVert}

\newcommand*\rfrac[2]{{}^{#1}\!/_{#2}}

\theoremstyle{remark}
\newtheorem{remark}{Remark} 
\theoremstyle{example}
\theoremstyle{assumption}
\IEEEoverridecommandlockouts
\definecolor{orange}{rgb}{0.8627,0.4314,0.1961}
\definecolor{grey}{rgb}{0.2196,0.2471,0.3176}
\definecolor{blue}{rgb}{0.0275,0.4431,0.5294}
\definecolor{green}{rgb}{0,0.6078,0.4392}
\begin{document}

	\title{Location-Aided Coordinated Analog Precoding \mbox{\!for Uplink Multi-User Millimeter Wave Systems}}
	\author{
     	\IEEEauthorblockN{
     		Flavio Maschietti\IEEEauthorrefmark{3}, 
     		David Gesbert\IEEEauthorrefmark{3}, 
     		Paul de Kerret\IEEEauthorrefmark{3}}
     	\\\IEEEauthorblockA{ 
     		\IEEEauthorrefmark{3}Communication Systems Department, EURECOM, Sophia-Antipolis, France\\
    		Email: \{flavio.maschietti, david.gesbert, paul.dekerret\}@eurecom.fr}
    }
	\maketitle
	\vspace{-0.5cm}
	\begin{abstract}
		
		Millimeter wave (mmWave) communication is expected to have an important role in next generation cellular networks,
		aiming to cope with the bandwidth shortage affecting conventional wireless carriers.
		Using side-information has been proposed as a potential approach to accelerate beam selection in mmWave massive MIMO 
		(m-MIMO) communications. However, in practice, such information is not error-free, leading to performance degradation.
		In the multi-user case, a wrong beam choice might result in irreducible inter-user interference at the base station (BS) side.
		In this paper, we consider location-aided precoder design in a mmWave uplink scenario with multiple users (UEs). 
		Assuming the existence of direct device-to-device (D2D) links, we propose a decentralized coordination mechanism for
		robust fast beam selection. The algorithm allows for improved treatment of interference at the BS side and in turn leads 
		to greater spectral efficiencies.		
			
	\end{abstract}
	
	\section{Introduction}
		
		The large bandwidths available at mmWave carrier frequencies are expected to help meet the throughput requirements
		for future mobile networks~\cite{7400949}. Since smaller wavelength signals are more prone to absorption, 
		mmWave communications require beamforming in order to guarantee appropriate link margins and 
		coverage~\cite{6834753, 7109864}.
		To this end, m-MIMO techniques~\cite{6798744} are envisioned as high-gain directional antennas with small form factor
		can be designed for mmWave usage~\cite{5723707}. However, configuring those massive antennas to operate with large 
		bandwidths entails an additional effort. The high cost and power consumption of the radio components impact on the 
		UEs and small BSs, thus limiting the practical implementation of a fully-digital beamforming architecture~\cite{7400949}. 
		Moreover, the large number of antennas at both ends of the radio links would require unfeasible CSI-training overhead 
		to design the precoders.
		
		One step towards simplification consists in replacing the fully-digital architecture with a \emph{hybrid} analog-digital
		one~\cite{6717211, 6847111, 7389996, 7579557, 7961152, 7913599}. In mixed analog-digital architectures, 
		a low-dimensional digital processor is concatenated with an RF analog beamformer, implemented through phase shifters. 
		Note that while the latter is sufficient to achieve a good part of the overall beamforming gain -- through beam steering
		towards desired spatial directions -- the digital stage is essential when processing multiple streams and users.
		
		Interestingly, existing works on hybrid architectures typically ignore multi-user interference issues in the analog domain
		and cope with them in the digital part. For instance, in \cite{7160780}, a procedure is proposed for the downlink 
		transmission, where the analog stage is intended to find the best beam directions for each UE (regardless of multi-user
		interference), while the digital one applies the conventional Zero-Forcing (ZF) beamformer on the resulting effective
		channel. The strength of this approach lies in the fact that it is possible to use the existing beam training algorithms for
		single-user links -- such as~\cite{5262295, 6600706, 7947217} -- in the analog stage. Such algorithms have been
		developed bearing in mind the need for fast link establishment in low-latency applications. Nevertheless, the reduced
		number of digital chains might not always allow to resolve the residual multi-user interference which remains after the
		analog beamforming stage. In particular, in a mmWave propagation scenario~\cite{6834753, 7109864}, multiple closely
		located UEs will likely share some common reflectors, causing an alignment of the main path's angles of arrival at the BS
		receiver and preventing it from resolving the interference, even at the digital decoding stage.
					
		To solve this problem, a principal idea consists in treating interference before it takes place, i.e. the UE side, as is done for
		example in \cite{7433949, 7925850}. Although showing significant performance advantages over the existing solutions,
		these works assume perfect CSIR for analog beamforming and single-antenna UEs, which might not be realistic in all
		mmWave contexts~\cite{7160780}.
		
		Rather, we are interested in statistically-driven analog beamforming at the UE TX side. In this paper, we point out that
		simple analog UE beam selection can be designed so as to enable the analog receive beam on the BS side to discriminate
		for interference. We propose to do this through the help of low-rate side-information at the UEs.
		Several works can be found in the mmWave literature, where side-information is exploited to improve performance
		without burdening overhead. Side-information can be obtained from various sources, such as automotive
		sensors~\cite{7786130}, UHF band~\cite{AGHeath}, GNSS~\cite{7536855}, or also past multipath fingerprints
		measurements~\cite{VaCSBH17}.
		
		We bring forward the idea that position-based side-information can be exploited in order to develop a coordination
		mechanism between the UEs, so that the interference at the BS side can be treated efficiently through both 
		the analog and digital parts of the receiver, as opposed to relying on the digital part alone. The main intuition is to use
		coordination to make sure the selected analog beams at the BS convey the full rank of multi-user channels towards the
		digital part to preserve invertibility.
		
		As in some previous work~\cite{BAMGdK}, we are interested in establishing a \emph{robust} form of coordination 
		which accounts for possible noise in the positioning information made available to the UEs. However, ~\cite{BAMGdK}
		targets a single-user scenario only. In the multi-user scenario, the lack of a real-time communication channel prevents 
		the UEs from exchanging instantaneous CSI. We consider, instead, the existence of a low-rate unidirectional D2D 
		channel, allowing communication of GPS-type data. In particular, we consider a hierarchical set-up where higher
		ranked UEs receive position information from lower ranked ones. The unidirectional aspect and the limitation to 
		position information exchange help keep the D2D overhead much lower than real-time D2D. 
		Our main contributions read as follows:
		\begin{itemize}
			\item{We formulate the problem of per-user analog precoding with side position information and recast it as a
			\emph{decentralized} beam selection problem.}
			\item{Our algorithm exploits the hierarchical structure of the information, in order to perform robust (with respect to 
			position data noise) interference mitigation at both analog and digital stages.}
			\item{Under the proposed method, the UEs coordinate to select beams which, while being suboptimal in terms of
			average power, help attain the full rank condition needed at the BS for interference suppression.}
		\end{itemize}
	
	\section{System Model}

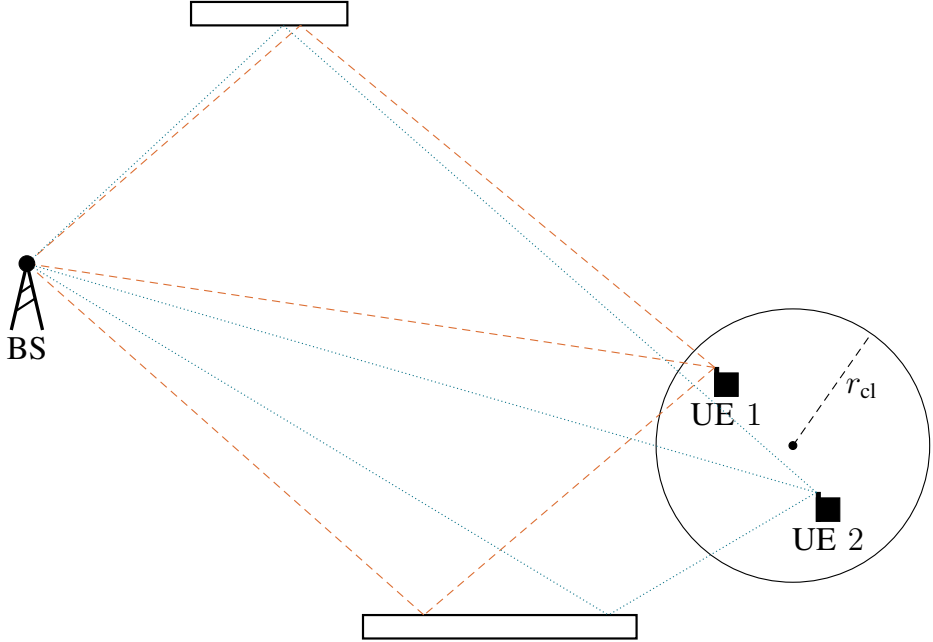
\begin{figure}[h]
			\centering
			\resizebox{12.73cm}{!}{
			\begin{tikzpicture}
				\draw (5,-1.18) circle (1.75cm);
				\filldraw[black] (5,-1.18) circle (.05cm);
				\draw[densely dashed] (5,-1.18) -- (6, .25);
				\draw (5.88,-.45) node {$r_{\text{cl}}$};
				\filldraw[black] (4,-.55) rectangle +(.3,.3);
				\filldraw[black] (4,-.55) rectangle +(.05,.37);
				\draw (4.15,-.78) node {UE $1$};
				\filldraw[black] (5.3,-2.15) rectangle +(.3,.3);
				\filldraw[black] (5.3,-2.15) rectangle +(.05,.37);
				\draw (5.45,-2.38) node {UE $2$};
				\draw[very thick] (-5,0.3) -- (-4.8,1.15);
				\draw[very thick] (-4.8,1.15) -- (-4.6,.3);
				\draw[thick] (-4.88,.78) -- (-4.73,.88);
				\draw[thick] (-4.93,.55) -- (-4.7,.7);
				\draw (-4.8, .07) node {BS};
				\draw[thick] (-2.7,4.2) rectangle +(2,.3);
				\draw[thick] (-.5,-3.35) rectangle +(3.5,-.3);
				\draw[orange, densely dashed] (4,-.18) -- (-1.3,4.2);
				\draw[orange, densely dashed] (-4.8, 1.15) -- (-1.3,4.2);
				\draw[orange, densely dashed] (4,-.18) -- (.28,-3.35);
				\draw[orange, densely dashed] (-4.8, 1.15) -- (.28,-3.35);
				\draw[orange, densely dashed] (4,-.18) -- (-4.8,1.15);
				\draw[blue, densely dotted] (5.3,-1.78) -- (-1.521,4.2);
				\draw[blue, densely dotted] (-4.8, 1.15) -- (-1.521,4.2);
				\draw[blue, densely dotted] (5.3,-1.78) -- (2.64,-3.35);
				\draw[blue, densely dotted] (-4.8, 1.15) -- (2.64,-3.35);
				\draw[blue, densely dotted] (5.3,-1.78) -- (-4.8,1.15);
				\filldraw[black] (-4.8,1.15) circle (.1cm); 
			\end{tikzpicture}
			}
			\caption{Scenario example with $L = 3$ propagation paths, two reflectors, and $K = 2$ UEs. 
			The UEs are assumed to reside in a disk of radius $r_{\text{cl}}$, which is relatively common in realistic scenarios, 
			e.g. dense UE distribution in a coffee house. In this illustration, two closely located UEs are sharing some reflectors, 
			and paths reflecting on the top reflector arrive quasi-aligned at the BS while originating from distinct UEs.}
			\label{fig:Scen}
		\end{figure}

		Consider the single-cell uplink multi-user mmWave scenario in Fig. \ref{fig:Scen}.
		The BS is equipped with $N_{\text{BS}} \gg 1$ antennas to support $K$ UEs with 
		$N_{\text{UE}} \gg 1$ antennas each. The UEs are assumed to reside in a disk of a given radius $r_{\text{cl}}$, 
		which will be used to control inter-UE average distance.
		Each UE sends one data stream to the BS. We assume that the BS has $N_{\text{RF}} = K$ RF chains available, 
		each one connected to all the $N_{\text{BS}}$ antennas, assuming a fully-connected hybrid architecture~\cite{7400949}.
		
		The $u$-th UE precodes the data $s^{u} \in \mathbb{C}$ through the analog precoding vector 
		$\mathbf{v}^{u} \in \mathbb{C}^{N_{\text{UE}} \times 1}$. We assume that the UEs have one RF chain each, 
		i.e. UEs are limited to analog beamforming via phase shifters (constant-magnitude elements)~\cite{6847111}. 
		In addition, $\mathbb{E}[\norm{\mathbf{v}^{u} s^{u}}^2] \le 1$, assuming normalized power constraints.
		
		The reconstructed signal after mixed analog/digital combining at the BS can be expressed as follows 
		-- assuming no timing and carrier mismatches:
		\begin{equation} \label{RX_Signal_pAnalogComb}
			\mathbf{\hat{x}} = \sum_{u=1}^K \mathbf{W}_{\text{D}} \mathbf{W}_{\text{RF}}^{\textrm{H}} 
			\mathbf{H}^{u} \mathbf{v}^{u} s^{u} + 
			\mathbf{W}_{\text{D}} \mathbf{W}_{\text{RF}}^{\textrm{H}} \mathbf{n}
		\end{equation}
		where $\mathbf{H}^{u} \in \mathbb{C}^{N_{\text{BS}} \times N_{\text{UE}}}$ is the channel matrix from the $u$-th UE 
		to the BS and $\mathbf{n} \in \mathbb{C}^{N_{\text{BS}} \times 1}$ is the thermal noise vector, with zero mean 
		and covariance matrix $\sigma_\mathbf{n}^2 \mathbf{I}_{N_{\text{BS}}}$. 
		$\mathbf{W}_{\text{RF}} \in \mathbb{C}^{N_{\text{BS}} \times N_{\text{RF}}}$ is, instead, the analog combining matrix,
		containing the vectors relative to each of the $K$ RF chains (subject to the same hardware constraints as described above), 
		while  $\mathbf{W}_{\text{D}} \in \mathbb{C}^{K \times N_{\text{RF}}}$ denotes the digital combining matrix.			
		
		The received SINR for the $u$-th UE at the BS is expressed as follows:
		\begin{equation} \label{SINR}
			\gamma^{u} = 
			\frac{|\mathbf{w}_{\text{D}}^{u} \mathbf{W}_{\text{RF}}^{\textrm{H}} \mathbf{H}^{u} \mathbf{v}^{u}|^2}
			{\sum_{w \ne u}|\mathbf{w}_{\text{D}}^{u} 
			\mathbf{W}_{\text{RF}}^{\textrm{H}} \mathbf{H}^{w} \mathbf{v}^{w}|^2 + \sigma_{\mathbf{\tilde{n}}}^2}
		\end{equation}
		where $\mathbf{w}_{\text{D}}^{u} \in \mathbb{C}^{1 \times N_{\text{RF}}}$ denotes the row of 
		$\mathbf{W}_{\text{D}}$ related to the $u$-th UE (one RF chain for each UE), and where we used the short-hand notation 
		$\mathbf{\tilde{n}} = \mathbf{W}_{\text{D}} \mathbf{W}_{\text{RF}}^{\textrm{H}} \mathbf{n}$ for the filtered 
		thermal noise.
		
		\subsection{Channel Model}
			
			Unlike the conventional UHF band propagation environment, the mmWave one does not exhibit 
			rich-scattering~\cite{6834753} and is in fact modeled as a geometric channel with a limited 
			number of dominant propagation paths which survive high attenuation. The UE $u$ is thus subject to the channel 
			matrix $\mathbf{H}^{u} \in \mathbb{C}^{N_{\textrm{BS}} \times N_{\textrm{UE}}}$, expressed as the sum of 
			$L$ components or contributions~\cite{6847111}:
			\begin{equation} \label{H}
				\mathbf{H}^{u} = \big(N_{\textrm{BS}} N_{\textrm{UE}}\big)^{\rfrac{1}{2}} \Big(
					\sum_{\ell=1}^L \alpha_{\ell}^{u} \mathbf{a}_{\textrm{BS}}(\vartheta_{\ell}^{u})
					\mathbf{a}^{\textrm{H}}_{\textrm{UE}} (\phi_{\ell}^{u})
				\Big)
			\end{equation}
			where $\alpha_{\ell}^{u} \sim \mathcal{CN}(0, (\sigma_\ell^{u})^2)$ denotes the complex gain for the $\ell$-th 
			path of the $u$-th UE. Furthermore, we assume that the variances 
			$(\sigma_\ell^{u})^2, \ell \in \{ 1, \dots, L \}; u \in \{ 1, \dots, K \}$ of the paths are such as 
			$\sum_\ell (\sigma_\ell^{u})^2 = 1, ~\forall u \in \{ 1, \dots, K \}$.
			
			The variables $\phi_\ell^{u} \in [0, 2\pi)$ and $\vartheta_\ell^{u} \in [0, 2\pi)$ are the angles of departure (AoDs) and 
			arrival (AoAs) for each contribution, for a given UE $u$, where one angle pair corresponds to the LoS direction while
			other might account for the presence of strong reflectors (e.g. buildings, hills) in the environment. 
			The positions of those points of reflection depend on the position of the considered UE (see Fig. \ref{fig:Scen}). 
			We will denote the reflecting points for the $u$-th UE with $\text{R}^u_i, i \in \{ 1, \dots, L-1 \}$ in the rest of the paper.
			
			The vectors $\mathbf{a}_{\textrm{UE}}(\phi_\ell^{u}) \in \mathbb{C}^{N_{\textrm{UE}} \times 1}$ and 
			$\mathbf{a}_{\textrm{BS}}(\vartheta_\ell^{u}) \in \mathbb{C}^{N_{\textrm{BS}} \times 1}$ denote the antenna
			steering vectors at the $u$-th UE and the BS for the corresponding AoDs $\phi_\ell^{u}$ and AoAs $\vartheta_\ell^{u}$,
			respectively. 
			
			We assume to use ULAs at both sides, so that~\cite{Tse:2005}:
			\begin{align}
				\mathbf{a}_{\text{UE}}(\phi) =
					\frac{1}{(N_{\text{UE}})^{\rfrac{1}{2}}}
					\Big[
					{1, e^{\!\!-i \pi \cos(\phi)}, \dots, e^{\!\!-i \pi (N_{\text{UE}} - 1) \cos(\phi)}}
					\Big]^{\!\mathrm{T}} \\
				\mathbf{a}_{\text{BS}}(\vartheta) =
					\frac{1}{(N_{\text{BS}})^{\rfrac{1}{2}}}
					\Big[
					{1, e^{\!\!-i \pi \cos(\vartheta)}, \dots, e^{\!\!-i \pi (N_{\text{BS}} - 1) \cos(\vartheta)}} 
					\Big]^{\!\mathrm{T}}
			\end{align}
		
		\subsection{Codebooks for Analog Beams}
		
			The most recognized method to implement the analog beamformer is through a network of \emph{digitally-controlled}
			phase shifters~\cite{5556449} (refer to \cite{7400949} for alternative architectures). 
			Thus, the phase of each element of the analog beamformer is limited to fixed quantized values, and therefore,
			the beamforming vectors need to be selected from a finite set (or codebook).
			We denote the codebooks used for analog beamforming as:
			\begin{equation}
				\mathcal{V}_{\textrm{UE}} = \{ \mathbf{v}_1, \dots, \mathbf{v}_{M_{\text{UE}}} \}, \quad
				\mathcal{V}_{\textrm{BS}} = \{ \mathbf{w}_1, \dots, \mathbf{w}_{M_{\text{BS}}} \}
			\end{equation}
			where $\mathcal{V}_{\textrm{UE}}$ is assumed to be shared between all the UEs, to ease the notation.
			
			For ULAs, a suitable design for the fixed beamforming vectors in the codebook consists in selecting steering vectors 
			over a discrete grid of angles~\cite{6600706, 7160780}:
			\begin{equation} \label{AnBeamform1}
				\mathbf{v}_p = \mathbf{a}_{\textrm{UE}}(\bar{\phi}_p), \quad p \in \{1, \dots, M_{\text{UE}}\}
			\end{equation}
			\begin{equation}  \label{AnBeamform2}
				\mathbf{w}_q = \mathbf{a}_{\text{BS}}(\bar{\vartheta}_q), \quad q \in \{1, \dots, M_{\text{BS}}\}
			\end{equation}
			where the angles $\bar{\phi}_p, ~\forall p \in \{ 1, \dots, M_{\text{UE}} \}$ and 
			$\bar{\vartheta}_q, ~\forall q \in \{ 1, \dots, M_{\text{BS}} \}$ can be chosen according to different strategies, 
			including regular and non-regular sampling of the $[0, \pi]$ range~\cite{BAMGdK}.
			\begin{remark}
				Given the one-to-one correspondence between the beamforming vectors in $\mathcal{V}_{\text{BS}}$
				(resp. $\mathcal{V}_{\text{UE}}$), and the grid angles $\bar{\vartheta}_q, ~\forall q \in \{ 1, \dots, M_{\text{BS}} \}$
				(resp. $\bar{\phi}_p, ~\forall p \in \{ 1, \dots, M_{\text{UE}} \}$), we will make the abuse of notation 
				$q \in \mathcal{V}_{\text{BS}}$ (resp. $p \in \mathcal{V}_{\text{UE}}$) to denote the vector
				$\mathbf{w}_q \in \mathcal{V}_{\text{BS}}$ (resp. $\mathbf{v}_p \in \mathcal{V}_{\text{UE}}$). \qed
			\end{remark}

\section{Information Model}
	
	In this section, we describe the structure of the channel state- and side-information available at both UEs and BS sides.
	We start with defining the nature of information in an ideal setting before turning to a realistic \emph{(noisy)} case.

	\begin{definition}
		The average beam gain matrix $\mathbf{G}^{u} \in \mathbb{R}^{M_{\textnormal{BS}} \times M_{\textnormal{UE}}}$
		contains the power level associated with each combined choice of analog beam pair between the BS and the $u$-th UE 
		after averaging over small scale fading. It is defined as:
		\begin{equation} \label{PerfectG}
			G_{q, p}^{u} = 
			\mathbb{E}_{\bm{\alpha}^{u}}\Big[ \big|\mathbf{w}^{\mathrm{H}}_q \mathbf{H}^{u} \mathbf{v}_p \big|^2\Big]
		\end{equation}
		where the expectation is carried out over the channel coefficients 
		$\bm{\alpha}^{u} = [\alpha^{u}_1, \alpha_2^{u}, \dots, \alpha_L^{u}]$ 
		and with $G_{q, p}^{u}$ denoting the $(q, p)$-element of $\mathbf{G}^{u}$.
	\end{definition}
	\begin{definition}
		The position matrix $\mathbf{P}^{u} \in \mathbb{R}^{2 \times (L+1)}$ contains the two-dimensional location
		coordinates $\mathbf{p}_n^{u} = [p_{n_x}^{u} \quad p_{n_y}^{u}]^\mathrm{T}$ for node $n$, where $n$ 
		indifferently refers to either the BS, the $u$-th UE or one of the reflectors $\textnormal{R}_i^{u}, i \in \{ 1, \dots, L-1 \}$. 
		It is defined as follows:
		\begin{equation}
			\mathbf{P}^{u} = 
			\begin{bmatrix} 
				\mathbf{p}^{u}_{\textnormal{BS}} &
				\mathbf{p}^{u}_{\textnormal{R$_1$}} &
				\dots &
				\mathbf{p}^{u}_{\textnormal{R$_{L-1}$}} &
				\mathbf{p}^{u}_{\textnormal{UE}} 
			\end{bmatrix}
		\end{equation}
		We will denote as $\mathcal{P}$ the set containing all the position matrices 
		$\mathbf{P}^u, ~\forall u \in \{ 1,\dots,K \}$.
	\end{definition}
	As shown in~\cite{BAMGdK}, the matrix $\mathbf{G}^{u}$ can be expressed as a function of the matrix $\mathbf{P}^{u}$. 
	We recall here the deterministic relationship that is found between those two matrices.
	\begin{lemma} \label{Lem1}
		We can write the average beam gain matrix relative to the $u$-th UE as follows:
		\begin{equation} \label{G_L_Functions}
			G_{q, p}^u(\mathbf{P}^u) = 
			\sum_{\ell=1}^L 
			(\sigma_\ell^u)^2 |L_{\textnormal{BS}}(\Delta_{\ell, q}^u)|^2 |L_{\textnormal{UE}}(\Delta_{\ell, p}^u)|^2
		\end{equation}
		where we remind the reader that $(\sigma_{\ell}^u)^2$ denotes the variance of the channel coefficients $\alpha_{\ell}^u$ 
		and we have defined:
		\begin{align} 
			\label{L_functions_1a}
			L_{\textnormal{UE}}(\Delta_{\ell, p}^u) &= \frac{1}{(N_{\textnormal{UE}})^{\rfrac{1}{2}}}
			\frac{e^{i (\pi/2) \Delta_{\ell, p}^u}}{e^{i (\pi/2) N_{\textnormal{UE}} \Delta_{\ell, p}^u}} 
			\frac{\sin((\pi/2) N_{\textnormal{UE}} \Delta_{\ell, p}^u)}{\sin((\pi/2) \Delta_{\ell, p}^u)} \\ 
			\label{L_functions_2a}
			L_{\textnormal{BS}}(\Delta_{\ell, q}^u) &= \frac{1}{(N_{\textnormal{BS}})^{\rfrac{1}{2}}}
			\frac{e^{i (\pi/2) \Delta_{\ell, q}^u}}{e^{i (\pi/2) N_{\textnormal{BS}} \Delta_{\ell, q}^u}} 
			\frac{\sin((\pi/2) N_{\textnormal{BS}} \Delta_{\ell, q}^u)}{\sin((\pi/2) \Delta_{\ell, q}^u)}
		\end{align}
		and
		\begin{equation}
			\Delta_{\ell, p}^u = (\cos(\bar{\phi}_p) - \cos(\phi_\ell^u))
		\end{equation}
		\begin{equation}
			\Delta_{\ell, q}^u = (\cos(\vartheta_\ell^u) - \cos(\bar{\vartheta}_q))
		\end{equation}
		with the angles $\phi_\ell^u, \ell \in \{ 1, \dots, L \}$ and $\vartheta_\ell^u, \ell \in \{ 1, \dots, L \}$ obtained from the 
		position matrix $\mathbf{P}^u$ through simple algebra (refer to~\cite{BAMGdK} for more details).
	\end{lemma}
	
	\subsection{Distributed Noisy Information Model}
		
		In the distributed model, each UE $u$ receives its own estimates of the position matrices 
		$\mathbf{P}^{w}, ~\forall w \in \{ 1, \dots, K \}$. We will use the superscript with parenthesis ${(u)}$ to denote any
		information known at the $u$-th UE. In particular, we denote as 
		$\mathbf{\hat{P}}^{w, (u)} \in \mathbb{R}^{2 \times (L+1)}, ~\forall w \in \{1,\dots,K\}$ the local
		information available at the $u$-th UE about the position matrix $\mathbf{P}^w$.
		This information is modeled as follows: 
		\begin{equation}
			\mathbf{\hat{P}}^{w, (u)} = \mathbf{P}^{w} + \mathbf{E}^{w, (u)} \qquad ~\forall w \in \{1,\dots,K\}
		\end{equation}
		where $\mathbf{E}^{w, (u)}$ denotes the following matrix:
		\begin{equation}
		\mathbf{E}^{w, (u)} = \begin{bmatrix} 
				\mathbf{e}^{w, (u)}_{\textnormal{BS}} &
				\mathbf{e}^{w, (u)}_{\textnormal{R$_1$}} &
				\dots &
				\mathbf{e}^{w, (u)}_{\textnormal{R$_{L-1}$}} &
				\mathbf{e}^{w, (u)}_{\textnormal{UE}}  &
			\end{bmatrix}
		\end{equation}
		containing the random position errors which the $u$-th UE made in estimating $\mathbf{p}^{w}_n$.
		Such error comes with an arbitrary, yet known, probability density function $f_{\mathbf{e}^{w, (u)}_n}$.
		
		\begin{definition}
			We will denote as $\mathcal{\hat{P}}^{(u)}$, where:
			\begin{equation}
				\mathcal{\hat{P}}^{(u)} = \{ \mathbf{\hat{P}}^{1, (u)}, \dots, \mathbf{\hat{P}}^{K, (u)} \}
			\end{equation}
			the overall local information available at the $u$-th UE containing all the estimated position matrices 
			$\mathbf{\hat{P}}^{w, (u)}, ~\forall w \in \{ 1,\dots,K \}$.			
		\end{definition}
	
	\subsection{Hierarchical Location-Information Exchange}
	
		The hierarchical (or nested) model is a sub-case of the distributed model in which the $u$-th UE has access to the estimates 
		of the UEs $u+1, \dots, K$. As we will see in the next section, this information structure enables some coordination for just
		half of the overhead needed in a conventional two-way exchange mechanism. One consequence in particular is that the 
		$u$-th UE is able to retrieve the beam decisions carried out at (the less informed) UEs $u+1, \dots, K$.
		
	\subsection{Additional Information}
	
		In what follows the number of dominant paths, and their average powers $(\sigma_\ell^{u})^2, \ell \in \{ 1, \dots, L \}; 
		u \in \{ 1, \dots, K \}$ are assumed to be known at each UE, based on prior averaged measurements. 
		Likewise, statistical distributions $f_{\mathbf{e}^{w, (u)}_n}, ~\forall u, w$ are supposed to be quasi-static and 
		as such are supposed to be available to each UE. In other words, the $u$-th UE is aware of the amount of 
		error in the position estimates which it and other UEs have to cope with.
			
\section{Multi-User Location-Aided Hybrid Precoding} \label{sec:MU-LA-HP}

	In order to maximize the received SNR $\gamma^u$ defined in \eqref{SINR} for each UE, the mutual optimization of both
	analog and digital components must be taken into account. A common approach consists in decoupling the design, as the
	analog beamformer can be optimized in terms of long-term channel statistics, whereas the digital one can be made 
	dependent on instantaneous information~\cite{7908940}.
	
	\subsection{Uncoordinated Beam Selection}
		
		We first review here the approach given in \cite{7160780}, where the authors proposed to design the analog beamformers
		to maximize the received power for each UE, neglecting multi-user interference.
		Once the analog beamformers are fixed at both UE and BS sides, the design of the digital beamformer at the BS follows the
		conventional MU-MIMO approach. In this respect, a common choice is to consider ZF combining. 
		Therefore, the digital beamforming matrix $\mathbf{W}_{\text{D}}$  is the pseudo-inverse of the effective channel matrix
		$\mathbf{\tilde{H}} \in \mathbb{C}^{N_{\text{RF}} \times K}$, which is defined as follows~\cite{Tse:2005}:
		\begin{equation}
			\mathbf{W}_{\text{D}} =
			\big(\mathbf{\tilde{H}}^{\textrm{H}} \mathbf{\tilde{H}} \big)^{-1} \mathbf{\tilde{H}}^{\textrm{H}}
		\end{equation}
		where
		\begin{equation}
			\mathbf{\tilde{H}} = \begin{bmatrix}
				\mathbf{W}_{\text{RF}}^{\text{H}} \mathbf{H}^{1} \mathbf{v}^{1} & 
				\mathbf{W}_{\text{RF}}^{\text{H}} \mathbf{H}^{2} \mathbf{v}^{2} & \dots &
				\mathbf{W}_{\text{RF}}^{\text{H}} \mathbf{H}^{K} \mathbf{v}^{K}
			\end{bmatrix}
		\mbox{, and with }
			\mathbf{W}_{\text{RF}} = \begin{bmatrix} 
				\mathbf{w}^{1} & \mathbf{w}^{2} & \dots & \mathbf{w}^{K}
			\end{bmatrix}. \nonumber
		\end{equation}
	
			When position and path average power information is available, the beam selection 
			$(q_{u}^{\text{un}} \in \mathcal{V}_{\text{BS}}, p_{u}^{\text{un}} \in \mathcal{V}_{\text{UE}})$ at the analog stage of 
			the algorithm proposed in~\cite{7160780} -- which we will denote as \emph{uncoordinated} (un) -- can be expressed 
			as follows:
		\begin{equation} \label{UN_Alg}
			(q_{u}^{\text{un}}, p_{u}^{\text{un}}) = 
			\argmax_{\substack{q_u \in \mathcal{V}_{\text{BS}} \\ p_u \in \mathcal{V}_{\text{UE}}}}
			\mathcal{R}^{u}\Big(\mathcal{\hat{P}}^{(u)}, q_u, p_u\Big), ~\forall u
		\end{equation}
		where we have defined the single-user rate $\mathcal{R}^{u}$ as~\cite{BAMGdK}:
		\begin{equation} \label{SU_Rate}
			\mathcal{R}^{u}(\mathcal{P}, q_u, p_u) = 
			\log_2 \Big(1+\frac{G^{u}_{q_u, p_u}(\mathbf{P}^{u})}{\sigma_{\mathbf{n}}^2}\Big) 
		\end{equation}
		
		Equation \eqref{UN_Alg} can be solved through direct search of the maximum in the matrix $\mathbf{G}^{u}$, 
		derived from $\mathcal{\hat{P}}^{(u)}$ through \eqref{G_L_Functions}. 
		
\begin{figure}[h]
			\centering
			\includegraphics[trim=4.15cm 8.1cm 4.15cm 8.5cm, width=0.73\columnwidth]
			{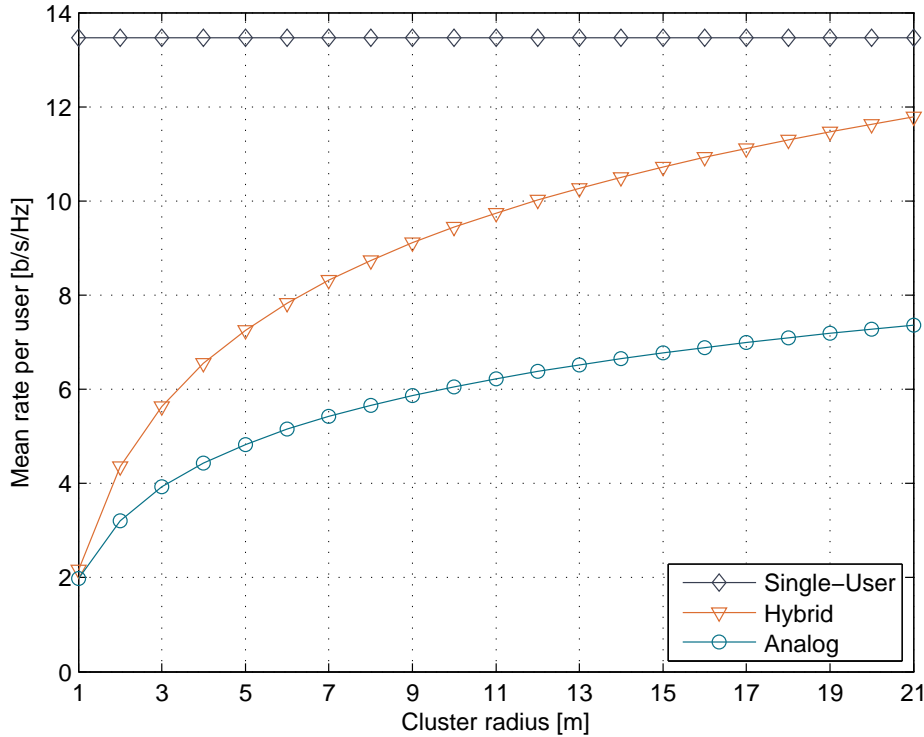}
			\caption{Mean rate per UE vs Cluster radius.
			The performance degrades \emph{sharply} as the inter-UE average distance decreases.}
			\label{fig:rate_vs_UEs_cluster_rad}
		\end{figure}

		While being simple to implement, the information at each UE in this method is treated as perfect, although some Bayesian
		robustization can be introduced~\cite{BAMGdK}. Another limitation of this approach is that each UE solves its own beam
		selection problem in a way which is independent of other UEs, thus ignoring the possible impairments in terms of
		interference. We illustrate this effect in Fig. \ref{fig:rate_vs_UEs_cluster_rad}, where we plot the mean rate per UE 
		obtained when the analog precoders are chosen through \eqref{UN_Alg}, in case of $K = 2$ UEs, \emph{perfect} position
		information, as a function of $r_{\text{cl}}$. 
		As the inter-UE average distance decreases, the performance of this procedure degrades, since the UEs have much more
		chance to share common best propagation paths (which results in severe interference at the analog stage at the BS).
		The action of the ZF is noticeable but not sufficient for small cluster radii. 
		In what follows, we consider different flavors of coordination.
	
	\subsection{Naive-Coordinated Beam Selection}
		
		In order to improve performance, we design the analog precoders according to the following figure of merit, which
		takes into account the average multi-user interference at the analog stage:
		\begin{equation} \label{MU_Rate}
			\mathcal{R}(\mathcal{P}, q_{1:K}, p_{1:K}) =
			\sum_{u=1}^K \log_2 \Big(1+\frac{G^{u}_{q_u, p_u}(\mathbf{P}^{u})}
			{\sum_w G_{q_u, p_w}^{w}(\mathbf{P}^{w}) + \sigma_{\mathbf{n}}^2}\Big)
		\end{equation}
		\begin{remark}
			We used here the short-hand $q_{1:K}, p_{1:K}$ to denote the indexes $q_1, \dots, q_K$ and $p_1, \dots, p_K$,
			respectively. \qed
		\end{remark}
		The hierarchical model allows the $u$-th UE to predict the beam selected at UEs $u+1, \dots, K$. 
		However, for a full coordination, the $u$-th UE would also need to know the precoding strategies of the
		\emph{more} informed UEs, i.e. UE $1, \dots, u-1$, which involves some guessing~\cite{1099850}.
		
		As a first approximation, the $u$-th UE can assume that its estimates are \emph{perfect} (error-free) and \emph{global}
		(shared between all the UEs). Since UEs $1, \dots, u-1$ have in fact different estimates, and since such information is not
		error-free, we call this approach \emph{naive-coordinated} (nc). The beam indexes
		$(q_{u}^{\text{nc}} \in \mathcal{V}_{\text{BS}}, p_{u}^{\text{nc}} \in \mathcal{V}_{\text{UE}})$ associated to the 
		$u$-th UE are then found as follows:
		\begin{equation} \label{Naive-Coord}
			(\tilde{q}_{1:{u-1}}, q_u^{\text{nc}},\tilde{p}_{1:{u-1}}, p_u^{\text{nc}}) =
			\argmax_{\substack{q_1, \dots, q_u \in \mathcal{V}_{\text{BS}} \\ p_1, \dots, p_u \in \mathcal{V}_{\text{UE}}}}
			\mathcal{R}_{q_{{u+1}:K}^o, p_{{u+1}:K}^o} 
			\Big(\mathcal{\hat{P}}^{(u)}, q_{1:u}, p_{1:u}\Big)
		\end{equation}
		\begin{remark}
			We make here an abuse of notation. The subscripts ${q_{{u+1}:K}^o, p_{{u+1}:K}^o}$
			acknowledge for the known strategies at the $u$-th UE. Those strategies are fixed parameters of the function
			$\mathcal{R}$. The same notation will be used in the rest of the paper. \qed
		\end{remark}
		\begin{remark}
			The $u$-th UE will use the precoding vector associated to the index $p_u^{\text{nc}} \in \mathcal{V}_{\text{UE}}$ to
			reach the BS and will discard the remaining beam indexes $\tilde{q}_{1:{u-1}}, \tilde{p}_{1:{u-1}}$ found for the other
			UEs. Indeed, those indexes only correspond to guesses realized at the $u$-th UE which do not necessarily correspond 
			to the true beams used for transmission at UEs $1, \dots, u-1$. We have introduced the notation $\tilde{q}_u$ to denote
			such beams. \qed
		\end{remark}
	
	\subsection{Statistically-Coordinated Beam Selection}
		
		The naive-coordinated approach relies on the correctness of the position estimates available at each UE. As a consequence,
		its performance is expected to degrade in case of GPS inaccuracies or lost location awareness. As the precision of position
		estimates decreases, the beam selection is expected to have more confidence in long-term statistics alone, to predict the 
		behavior of the UE which are higher ranked in the information chain. In this case, the position estimates are not exploited,
		and each UE relies on prior statistics to figure out other UEs' information.
		We denote the resulting \emph{statistically-coordinated} (sc) beam indexes relative to the $u$-th UE as 
		$(q_{u}^{\text{sc}} \in \mathcal{V}_{\text{BS}}, p_{u}^{\text{sc}} \in \mathcal{V}_{\text{UE}})$, which read as follows:
		\begin{equation} \label{Stat-Coord}
			(\tilde{q}_{1:{u-1}},q_u^{\text{sc}},\tilde{p}_{1:{u-1}}, p_u^{\text{sc}}) =
			\argmax_{\substack{q_1, \dots, q_u \in \mathcal{V}_{\text{BS}} \\ p_1, \dots, p_u \in \mathcal{V}_{\text{UE}}}}
			\mathbb{E}_{\mathcal{P}|r_{\text{cl}}} \Big[ 
			\mathcal{R}_{q_{{u+1}:K}^o, p_{{u+1}:K}^o}
			\Big(\mathcal{P}, q_{1:u}, p_{1:u}\Big)
			\Big]
		\end{equation}
		This is a long-term optimization which is updated only if prior statistics change.
		Thus, \eqref{Stat-Coord} represents a simple stochastic optimization problem~\cite{Shapiro} which can be solved 
		through e.g. approximation of the expectation operator (carried out over prior statistics) with Monte-Carlo iterations.
	
	\subsection{Robust-Coordinated Beam Selection}
		
		The UEs have also access to the statistics of their local position estimates. In the previous approach, each UE used prior
		statistics to guess the precoding strategies of the more informed UEs. This helps in case the local information available
		at the $u$-th UE is not accurate enough to gain more knowledge about the UEs $1, \dots, u-1$. In the opposite case, local
		statistical information can be exploited to supplement prior information. In this approach, the UEs look for beam selection
		strategies which \emph{progressively} pass from exploiting local information only -- in case of perfect local information -- 
		to exploiting statistical information only -- in case of poor local information. 
		We denote this approach as \emph{robust-coordinated} (rc). 
		The beams $(q_{u}^{\text{rc}} \in \mathcal{V}_{\text{BS}}, p_{u}^{\text{rc}} \in \mathcal{V}_{\text{UE}})$ for the 
		$u$-th UE are obtained through:
		\begin{equation} \label{Robust-Coord}
			(\tilde{q}_{1:{u-1}},q_u^{\text{rc}},\tilde{p}_{1:{u-1}}, p_u^{\text{rc}}) =
			\argmax_{\substack{q_1, \dots, q_u \in \mathcal{V}_{\text{BS}} \\ p_1, \dots, p_u \in \mathcal{V}_{\text{UE}}}}
			\mathbb{E}_{\mathcal{P}|\mathcal{\hat{P}}^{(u)}, r_{\text{cl}}} \Big[ 
			\mathcal{R}_{q_{{u+1}:K}^o, p_{{u+1}:K}^o}
			\Big(\mathcal{P}, q_{1:u}, p_{1:u}\Big)
			\Big]		
		\end{equation}
		\begin{remark}
			Here, the $u$-th UE considers its locally-available position estimates as \emph{imperfect} and \emph{globally}-shared.
			\qed
		\end{remark}
		Also in this case, an approximate solution can be obtained through Monte-Carlo methods, generating possible matrices
		$\mathcal{P}$ according to the (known) distribution $\mathcal{P}|\mathcal{\hat{P}}^{(u)}, r_{\text{cl}}$. 
		
		We summarize the proposed robust-coordinated beam selection used at the $u$-th UE in Algorithm \ref{R_PseudCod}. 
		In Step 1, the $u$-th UE retrieves the processing carried out at less informed UEs $u+1, \dots, K$. The $K$-th UE skips 
		this step. In Step 2, beam selection is performed through \eqref{MU_Rate} (an approximation) and \eqref{Robust-Coord}.
		
		\begin{algorithm} 
			\caption{$f_{\text{rc}}$ : Robust-Coordinated Beam Selection ($u$-th UE)}
			\label{R_PseudCod}
			\begin{algorithmic}[1]
				\small
				\algrenewcommand\algorithmicindent{2.0em}
				\Statex INPUT: 
				$\mathcal{\hat{P}}^{(w)}, ~\forall w \in \{u, \dots, K \}$, 
				pdf of $(\mathcal{P}|\mathcal{\hat{P}}^{(w)}, r_{\text{cl}}), ~\forall w \in \{u, \dots, K \}$
				\Statex{\textbf{Step 1}}
					\For {$w = K:u+1$} \Comment {The $K$-th UE skips this decreasing \emph{for} loop}
						\State $(q_w^o, p_w^o) = f_{\text{rc}}(\mathcal{\hat{P}}^{(w)}, q_{w+1:K}^o, p_{w+1:K}^o)$
					\EndFor
				\Statex{\textbf{Step 2}} 
					\State \textbf{return} $(q_u^{\text{rc}}, p_u^{\text{rc}}) \leftarrow$ Evaluate \eqref{Robust-Coord}
					\Comment {Refer to Algorithm \ref{R_PseudCod_DET} for implementation details}
			\end{algorithmic}
		\end{algorithm}
		\begin{algorithm} 
			\caption{Implementation details for Step 2 in Algorithm \ref{R_PseudCod} ($u$-th UE)} \label{R_PseudCod_DET}
			\begin{algorithmic}[1]
				\small
				\algrenewcommand\algorithmicindent{2.0em}
				\Statex INPUT: 
				$\mathcal{\hat{P}}^{(u)}$, pdf of $(\mathcal{P}|\mathcal{\hat{P}}^{(u)}, r_{\text{cl}})$,
				$q_{u+1}^o, \dots, q_{K}^o$, $p_{u+1}^o, \dots, p_{K}^o$
				\For {$i = 1:M$} \Comment {Approximate expectation over 
				$(\mathcal{P}|\mathcal{\hat{P}}^{(w)}, r_{\text{cl}}), ~\forall w$ with $M$ Monte-Carlo iterations}
					\State Generate possible position matrices 
					$\mathcal{P}$ through sampling over the distribution $(\mathcal{P}|\mathcal{\hat{P}}^{(w)}, r_{\text{cl}})$
					\State Compute possible gain matrices $\mathbf{\hat{G}}^{w}, ~\forall w \in \{1, \dots, K \}$ 
					through \eqref{G_L_Functions} using the generated $\mathcal{P}$
					\State \Call{SumRateEval}{$\mathbf{\hat{G}}^{1}, \dots, \mathbf{\hat{G}}^{K}, 
					q_{u+1}^o, \dots, q_{K}^o, p_{u+1}^o, \dots, p_{K}^o$}
					\Comment {Described in Algorithm \ref{R_PseudCod_Fun}}
				\EndFor
				\State Compute the average sum-rate over the Monte-Carlo iterations for all possible beam pairs
				\State $(q_u^{\text{rc}}, p_u^{\text{rc}}) \leftarrow$ 
				Indexes relative to the beams achieving maximum average sum-rate
				\State The pair of vectors with indexes $(q_u^{\text{rc}}, p_u^{\text{rc}})$ is assigned to the $u$-th UE
			\end{algorithmic}
		\end{algorithm}
		\begin{algorithm} 
			\caption{Function evaluating an approximated average sum-rate \eqref{MU_Rate} ($u$-th UE)} \label{R_PseudCod_Fun}
			\begin{algorithmic}[1]
				\small
				\Function {SumRateEval}{$\mathbf{\hat{G}}^{1}, \dots, \mathbf{\hat{G}}^{K}, 
					q_{u+1}^o, \dots, q_{K}^o, p_{u+1}^o, \dots, p_{K}^o$}
					\State $\textrm{KnownInds} = \{ q_{u+1}^o, \dots, q_{K}^o, p_{u+1}^o, \dots, p_{K}^o \}$
					\For {$w = 1:u-1$} \Comment {The most informed UE $1$ skips this loop}
						\State $(q_w, \dots, q_u, p_w, \dots, p_u) = 
						\max\big(\hat{G}_{q_w, p_w}^{w} / (\sum_{v \neq w} \hat{G}_{q_w, p_v}^{v} + N_0)\big)$
						\Comment {with given $\textrm{KnownInds}$ }
						\State $\textrm{KnownInds} = \{ q_w, p_w \} \cup \textrm{KnownInds}$
						\Comment {Updated set of indexes to be used in the next iteration}
						\State Discard all the other indexes $q_{w+1}, \dots, q_u, p_{w+1}, \dots, p_u$
					\EndFor
					\State \textbf{return} $\big(\hat{G}_{q_u, p_u}^{u} / (\sum_{w \neq u} \hat{G}_{q_u, p_w}^{w} + N_0)\big)$
					for all possible $q_u, p_u$
					\Comment {with given KnownInds}
				\EndFunction
			\end{algorithmic}
		\end{algorithm}
	
\section{Simulation Results}

	We evaluate here the performance of the proposed algorithms. We consider $L = 3$ multipath components. 
	A distance of $100$ m is assumed from the UEs' cluster center and the BS. 
	The radius of the UEs' cluster is set to $r_{\text{cl}} = 7$ m. Both the BS and UEs are equipped with 
	$N_{\text{UE}} = N_{\text{BS}} = 64$ antennas (ULA). The number of elements in the beam codebooks is 
	$M_{\text{UE}} = M_{\text{BS}} = 64$, with grid angles spaced according to the inverse cosine function so as to 
	guarantee equal gain losses among adjacent angles~\cite{BAMGdK}. 
	All the plotted rates are the averaged -- over 10000 Monte-Carlo runs -- rates per UE.

	\subsection{Location Information Model}
	
		In the simulations, we adopt a uniform bounded error model for location information~\cite{7536855, BAMGdK}.
		In particular, we assume that all the position estimates lie somewhere inside disks centered in the actual positions
		$\mathbf{p}^{u}_n, n \in \{ \text{UE}, \text{BS}, \text{R}_i^u \}; i \in \{1, \dots, L-1\}; u \in \{1, \dots, K\}$.
		Let $S(r)$ be the two-dimensional closed ball centered at the origin and of radius $r$, which is 
		$S(r) = \{ \boldsymbol{\upsilon} \in \mathbb{R}^2 : \norm{\boldsymbol{\upsilon}} \le r \}$. We model the estimation 
		errors $\mathbf{e}_n^{w, (u)}$ as a random variable \emph{uniformly} distributed in $S(r_n^{w, (u)})$, where
		$r_n^{w, (u)}$ is the maximum positioning error for the node $n$ of the $w$-th UE as seen from the $u$-th UE.
	
	\subsection{Results and Discussion}
	
		To evaluate the performance, we start with a simple configuration with $K = 2$ UEs.
		
		\subsubsection{Strong LoS}

			In what follows, we consider a stronger (on average) LoS path with respect to the reflected paths, being indeed the 
			prominent propagation driver in mmWave bands~\cite{6834753, 7109864}. The reflected paths are assumed to 
			have the same average power. The average power of such paths is assumed to be shared across the UEs, 
			i.e. $(\sigma_\ell^{(1)})^2, = (\sigma_\ell^{(2)})^2, ~\forall \ell \in \{1, \dots, L\}$.
			
\begin{figure}[h]
				\centering
				\captionsetup[subfigure]{justification=centering}
				\begin{subfigure}[b]{0.48\columnwidth}
					\includegraphics[trim=4.15cm 8.1cm 4.15cm 8.35cm, width=\columnwidth]
					{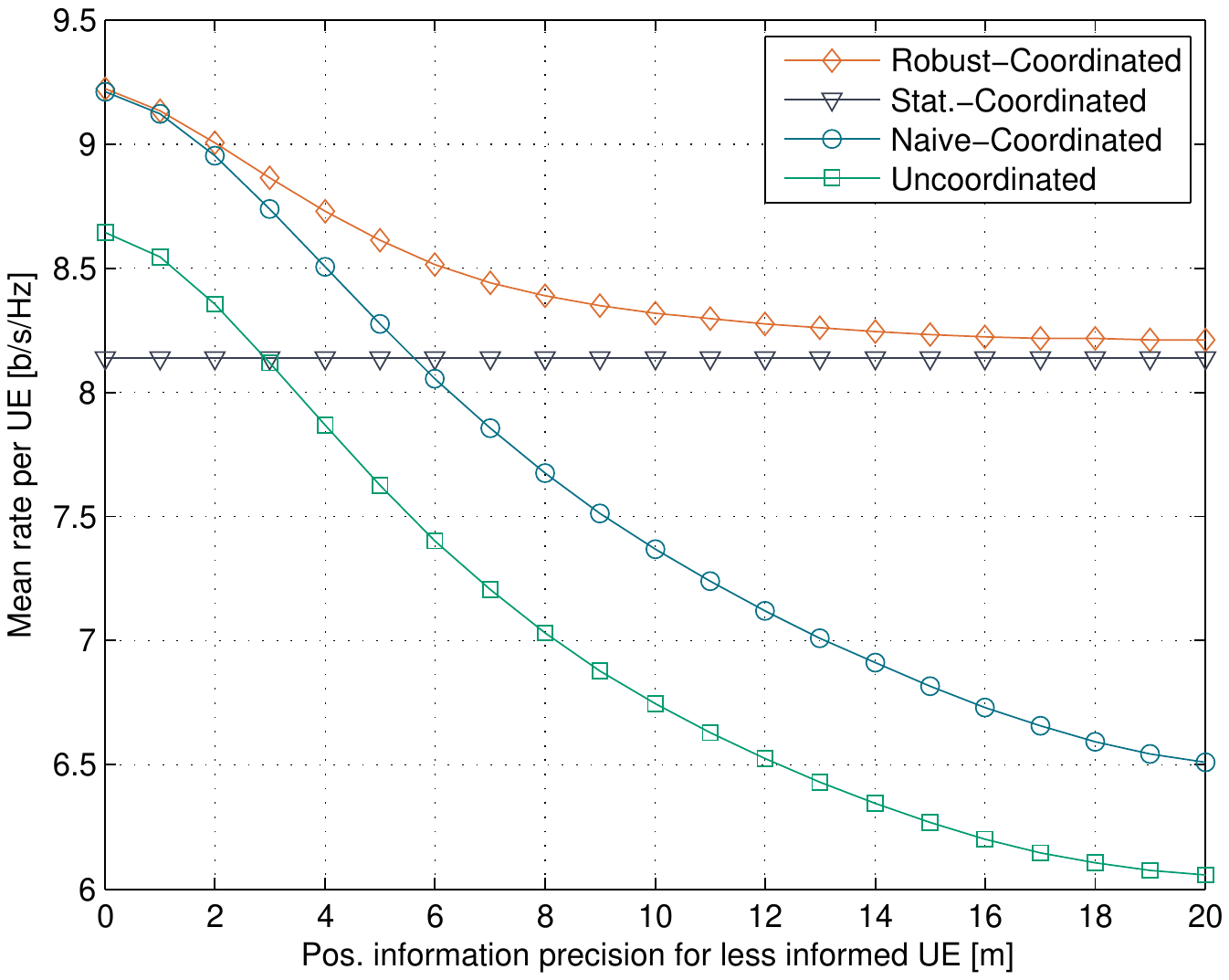}
					\caption{Most informed UE with \emph{perfect} information}
					\label{fig:rate_vs_error_rad_less_info_LoS_0}
				\end{subfigure}
				\quad
				\begin{subfigure}[b]{0.48\columnwidth}
					\includegraphics[trim=4.15cm 8.1cm 4.15cm 8.35cm, width=\columnwidth]
					{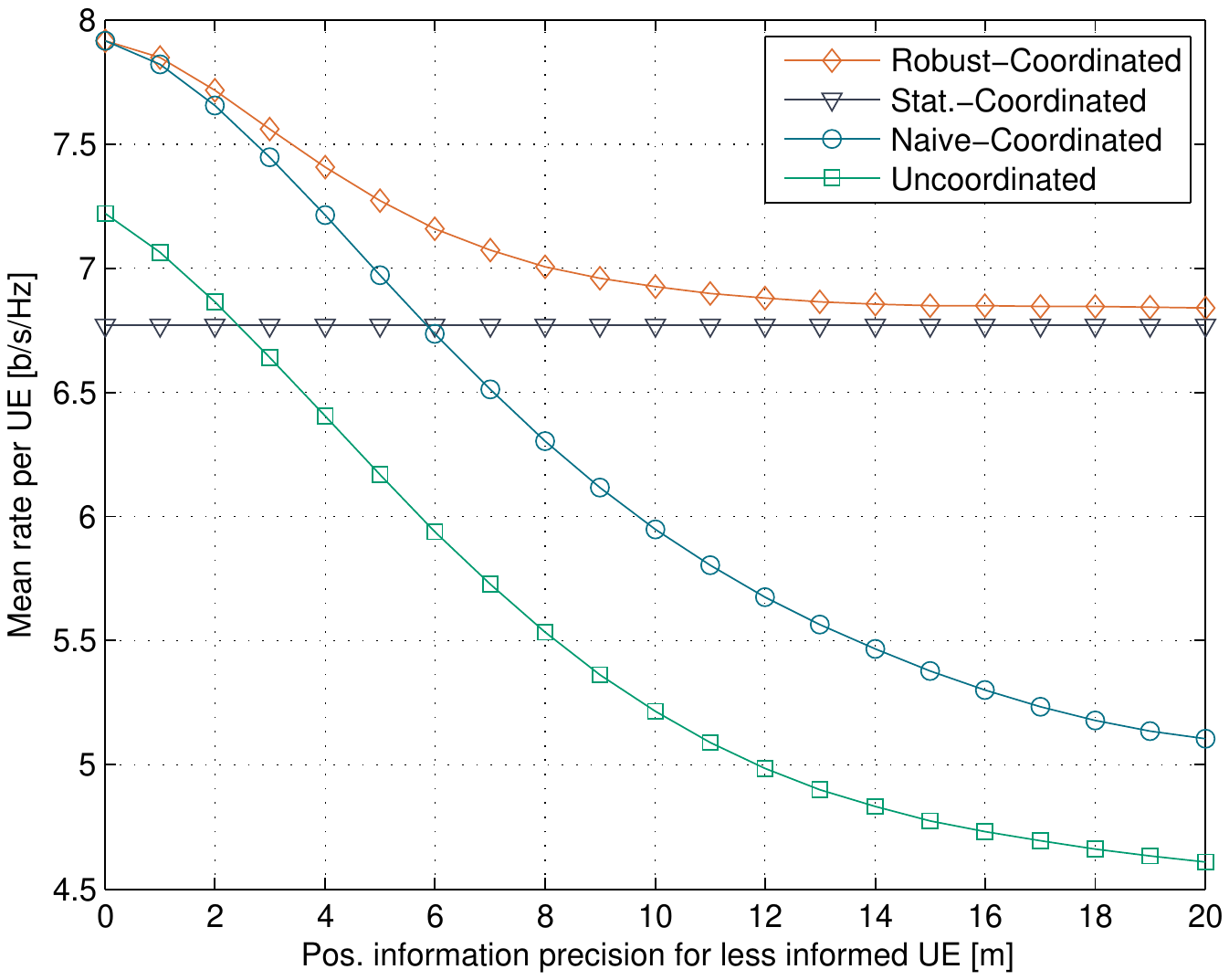}
					\caption{Most informed UE with $3$ m precision}
					\label{fig:rate_vs_error_rad_less_info_LoS_3}
				\end{subfigure}
				\caption{Mean rate per UE vs Position information precision. Strong LoS.}
				\label{fig:rate_vs_error_rad_less_info_LoS}
			\end{figure}

			In Fig. \ref{fig:rate_vs_error_rad_less_info_LoS}, we consider the performance of the proposed algorithm as a function 
			of the precision of the information available at the less informed UE.  In particular, an error radius of $5$ m means
			$r_n^{w, (2)} = 5$ m, $~\forall w, n$. As for the most informed UE, we consider \emph{perfect} information in Fig.
			\ref{fig:rate_vs_error_rad_less_info_LoS_0}, i.e. $r_n^{w, (1)} = 0$ m, $~\forall w, n$, and 3 m of precision in Fig. 
			\ref{fig:rate_vs_error_rad_less_info_LoS_3}, i.e. $r_n^{w, (1)} = 3$ m, $~\forall w, n$.
			
			Fig. \ref{fig:rate_vs_error_rad_less_info_LoS_0} and Fig. \ref{fig:rate_vs_error_rad_less_info_LoS_3} show that both
			the uncoordinated and the naive approaches degrade fast as the error radius for the less informed UE increases. 
			This is due to the fact that the UEs build their strategies according to their available position estimates, which
			become unreliable to perform beam selection. In particular, when the precision is less than $6$ m, 
			the statistically-coordinated approach -- based on statistical information only -- behaves better. 
			The robust-coordinated approach outperforms all the other algorithms, being able to discriminate for interference at
			the BS side, while taking into account the noise present in position information.
			
				
		\subsubsection{LoS Blockage}
		
			It is also interesting to observe how the proposed algorithms behave in case of total line-of-sight blockage, while having
			one stronger reflected path.
			
\begin{figure}[h]
				\centering
				\includegraphics[trim=4.15cm 8.1cm 4.15cm 8.5cm, width=0.81\columnwidth]
				{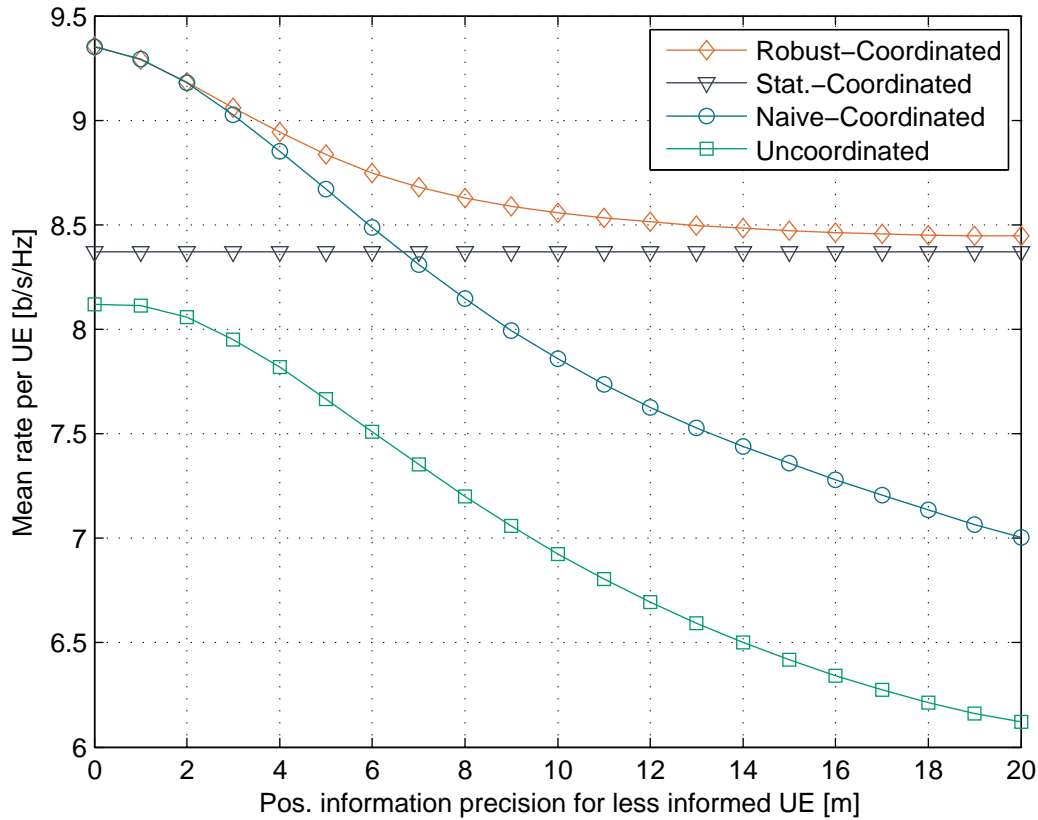}
				\caption{Mean rate per UE vs Position information precision. LoS blockage.}
				\label{fig:rate_vs_error_radius_less_informed_NLoS}
			\end{figure}

			Fig. \ref{fig:rate_vs_error_radius_less_informed_NLoS} compares the proposed algorithms as a function of the error
			radius for the less informed UE. The most informed UE is assumed to have access to \emph{perfect} information.
			The same considerations outlined for the strong LoS case remain valid. It is possible to observe that the uncoordinated
			approach performs worse than before (with strong LoS), since the UEs choose to use the beams pointing towards
			the same stronger reflected path. As a consequence, there is much more chance to arrive at the BS with
			non-distinguishable AoAs. In such cases, coordination between the UEs is essential to combat multi-user interference.
			
\section{Conclusions}

	In mmWave communications, multi-user interference has to be handled in the analog stage as well. 
	In this respect, suitable strategies for multi-user interference minimization can be applied in the beam
	domain through e.g. exploitation of location-dependent information.
	
	Dealing with the imperfections in location information is not trivial, due to the decentralized nature of the information, 
	which leads to disagreements between the UEs affecting performance. 
	In this work, we introduced a decentralized robust algorithm which aim to select the best precoder for each UE taking both 
	the noise present in location information and multi-user interference in the analog stage into account.
	
	Numerical experiments have shown that good performance can be achieved with the proposed algorithm and have confirmed
	that coordination is essential to counteract inter-UE interference in mmWave multi-user environments.
	
	Exploiting Machine Learning tools~\cite{Rasmussen2005} for solving the proposed algorithms in a more efficient manner is 
	an interesting and challenging problem which we aim to tackle in future studies.
	
\section*{Acknowledgments}
	
	F. Maschietti, D. Gesbert and P. de Kerret are supported by the ERC under the European Union's Horizon 2020 research 
	and innovation program (Agreement no. 670896).
			 
\bibliography{Bibl}

\begin{thebibliography}{10}
\providecommand{\url}[1]{#1}
\csname url@samestyle\endcsname
\providecommand{\newblock}{\relax}
\providecommand{\bibinfo}[2]{#2}
\providecommand{\BIBentrySTDinterwordspacing}{\spaceskip=0pt\relax}
\providecommand{\BIBentryALTinterwordstretchfactor}{4}
\providecommand{\BIBentryALTinterwordspacing}{\spaceskip=\fontdimen2\font plus
\BIBentryALTinterwordstretchfactor\fontdimen3\font minus
  \fontdimen4\font\relax}
\providecommand{\BIBforeignlanguage}[2]{{%
\expandafter\ifx\csname l@#1\endcsname\relax
\typeout{** WARNING: IEEEtran.bst: No hyphenation pattern has been}%
\typeout{** loaded for the language `#1'. Using the pattern for}%
\typeout{** the default language instead.}%
\else
\language=\csname l@#1\endcsname
\fi
#2}}
\providecommand{\BIBdecl}{\relax}
\BIBdecl

\bibitem{7400949}
R.~W. Heath, N.~Gonz\'alez-Prelcic, S.~Rangan, W.~Roh, and A.~M. Sayeed, ``An
  overview of signal processing techniques for millimeter wave {MIMO}
  systems,'' \emph{IEEE J. Sel. Topics Signal Process.}, Apr. 2016.

\bibitem{6834753}
M.~R. Akdeniz, Y.~Liu, M.~K. Samimi, S.~Sun, S.~Rangan, T.~S. Rappaport, and
  E.~Erkip, ``Millimeter wave channel modeling and cellular capacity
  evaluation,'' \emph{IEEE J. Sel. Areas Commun.}, June 2014.

\bibitem{7109864}
T.~S. Rappaport, G.~R. MacCartney, M.~K. Samimi, and S.~Sun, ``Wideband
  millimeter-wave propagation measurements and channel models for future
  wireless communication system design,'' \emph{IEEE Trans. Commun.}, Sept.
  2015.

\bibitem{6798744}
L.~Lu, G.~Y. Li, A.~L. Swindlehurst, A.~Ashikhmin, and R.~Zhang, ``An overview
  of massive {MIMO}: Benefits and challenges,'' \emph{IEEE J. Sel. Topics
  Signal Process.}, Oct. 2014.

\bibitem{5723707}
B.~Biglarbegian, M.~Fakharzadeh, D.~Busuioc, M.~R. Nezhad-Ahmadi, and
  S.~Safavi-Naeini, ``Optimized microstrip antenna arrays for emerging
  millimeter-wave wireless applications,'' \emph{IEEE Trans. Antennas Propag.},
  May 2011.

\bibitem{6717211}
O.~E. Ayach, S.~Rajagopal, S.~Abu-Surra, Z.~Pi, and R.~W. Heath, ``Spatially
  sparse precoding in millimeter wave {MIMO} systems,'' \emph{IEEE Trans.
  Wireless Commun.}, Mar. 2014.

\bibitem{6847111}
A.~Alkhateeb, O.~E. Ayach, G.~Leus, and R.~W. Heath, ``Channel estimation and
  hybrid precoding for millimeter wave cellular systems,'' \emph{IEEE J. Sel.
  Topics Signal Process.}, Oct. 2014.

\bibitem{7389996}
F.~Sohrabi and W.~Yu, ``Hybrid digital and analog beamforming design for
  large-scale antenna arrays,'' \emph{IEEE J. Sel. Topics Signal Process.},
  Apr. 2016.

\bibitem{7579557}
C.~Rusu, R.~M\`endez-Rial, N.~Gonz\'alez-Prelcic, and R.~W. Heath, ``Low
  complexity hybrid precoding strategies for millimeter wave communication
  systems,'' \emph{IEEE Trans. Wireless Commun.}, Dec. 2016.

\bibitem{7961152}
K.~Venugopal, A.~Alkhateeb, N.~Gonz\'alez-Prelcic, and R.~W. Heath, ``Channel
  estimation for hybrid architecture-based wideband millimeter wave systems,''
  \emph{IEEE J. Sel. Areas Commun.}, Sept. 2017.

\bibitem{7913599}
F.~Sohrabi and W.~Yu, ``Hybrid analog and digital beamforming for {mmWave}
  {OFDM} large-scale antenna arrays,'' \emph{IEEE J. Sel. Areas Commun.}, Jul.
  2017.

\bibitem{7160780}
A.~Alkhateeb, G.~Leus, and R.~W. Heath, ``Limited feedback hybrid precoding for
  multi-user millimeter wave systems,'' \emph{IEEE Trans. Wireless Commun.},
  Nov. 2015.

\bibitem{5262295}
J.~Wang, ``{Beam codebook based beamforming protocol for multi-Gbps
  millimeter-wave WPAN systems},'' \emph{IEEE J. Sel. Areas Commun.}, Oct.
  2009.

\bibitem{6600706}
S.~Hur, T.~Kim, D.~J. Love, J.~V. Krogmeier, T.~A. Thomas, and A.~Ghosh,
  ``Millimeter wave beamforming for wireless backhaul and access in small cell
  networks,'' \emph{IEEE Trans. Commun.}, Oct. 2013.

\bibitem{7947217}
D.~Ogbe, D.~J. Love, and V.~Raghavan, ``Noisy beam alignment techniques for
  reciprocal {MIMO} channels,'' \emph{IEEE Trans. Signal Process.}, Oct. 2017.

\bibitem{7433949}
J.~Li, L.~Xiao, X.~Xu, and S.~Zhou, ``Robust and low complexity hybrid
  beamforming for uplink multiuser {mmWave} {MIMO} systems,'' \emph{IEEE
  Commun. Lett.}, June 2016.

\bibitem{7925850}
Y.~Zhu and T.~Yang, ``Low complexity hybrid beamforming for uplink multiuser
  {mmWave} {MIMO} systems,'' in \emph{Proc. IEEE Wireless Commun. and Netw.
  Conf. (WCNC)}, Mar. 2017.

\bibitem{7786130}
J.~Choi, V.~Va, N.~Gonz\'alez-Prelcic, R.~Daniels, C.~R. Bhat, and R.~W. Heath,
  ``Millimeter-wave vehicular communication to support massive automotive
  sensing,'' \emph{IEEE Commun. Mag.}, Dec. 2016.

\bibitem{AGHeath}
\BIBentryALTinterwordspacing
A.~Ali, N.~Gonz\'alez-Prelcic, and R.~W. Heath, ``Millimeter wave
  beam-selection using out-of-band spatial information,'' \emph{CoRR}, Feb.
  2017. [Online]. Available: \url{http://arxiv.org/abs/1702.08574}
\BIBentrySTDinterwordspacing

\bibitem{7536855}
N.~Garcia, H.~Wymeersch, E.~G. Str\"om, and D.~Slock, ``{Location-aided mm-wave
  channel estimation for vehicular communication},'' in \emph{Proc. IEEE Int.
  Workshop on Signal Process. Advances in Wireless Commun. (SPAWC)}, Jul. 2016.

\bibitem{VaCSBH17}
\BIBentryALTinterwordspacing
V.~Va, J.~Choi, T.~Shimizu, G.~Bansal, and R.~W.~H. Jr., ``Inverse multipath
  fingerprinting for millimeter wave {V2I} beam alignment,'' \emph{CoRR}, 2017.
  [Online]. Available: \url{http://arxiv.org/abs/1705.05942}
\BIBentrySTDinterwordspacing

\bibitem{BAMGdK}
F.~Maschietti, D.~Gesbert, P.~de~Kerret, and H.~Wymeersch, ``Robust
  location-aided beam alignment in millimeter wave massive {MIMO},''
  \emph{Proc. IEEE Global Telecommun. Conf. (GLOBECOM)}, Dec. 2017.

\bibitem{Tse:2005}
D.~Tse and P.~Viswanath, \emph{Fundamentals of Wireless Communication}.\hskip
  1em plus 0.5em minus 0.4em\relax Cambridge University Press, 2005.

\bibitem{5556449}
Y.~Yu,  \emph{et~al.}, ``A 60 {GHz} phase shifter integrated with {LNA} and
  {PA} in 65 nm {CMOS} for phased array systems,'' \emph{IEEE J. Solid-State
  Circuits}, Sept. 2010.

\bibitem{7908940}
Z.~Li, S.~Han, and A.~F. Molisch, ``Optimizing channel-statistics-based analog
  beamforming for millimeter-wave multi-user massive {MIMO} downlink,''
  \emph{IEEE Trans. Wireless Commun.}, Jul. 2017.

\bibitem{1099850}
Y.-C. Ho and K.-C. Chu, ``{Team decision theory and information structures in
  optimal control problems},'' \emph{IEEE Trans. Autom. Control}, Feb. 1972.

\bibitem{Shapiro}
A.~Shapiro, D.~Dentcheva, and A.~Ruszczy\'nski, \emph{Lectures on stochastic
  programming: modeling and theory}.\hskip 1em plus 0.5em minus 0.4em\relax
  Philadelphia, PA, USA: Society for Industrial and Applied Mathematics, 2014.

\bibitem{Rasmussen2005}
C.~Rasmussen and C.~Williams, \emph{Gaussian Processes for Machine Learning},
  ser. Adaptive Computation and Machine Learning.\hskip 1em plus 0.5em minus
  0.4em\relax Cambridge, MA, USA: The MIT Press, 2006.

\end{thebibliography}
\bibliographystyle{IEEEtran}
			
\end{document}